# Witness electron beam injection using an active plasma lens for a proton beam-driven plasma wakefield accelerator

S.-Y. Kim, K. Moon, and M. Chung[*]

*Department of Physics, Ulsan National Institute of Science and Technology, Ulsan 44919, Republic of Korea*

K. N. Sjobak and E. Adli

*Department of Physics, University of Oslo, 0316 Oslo, Norway*

S. Doebert[†]

*BE Department, CERN, Geneva, CH-1211, Switzerland*

M. Dayyani

*Institute for Research in Fundamental Sciences, 19395-5531, Tehran, Iran*

E. S. Yoon

*Department of Nuclear Engineering, Ulsan National Institute of Science and Technology, Ulsan 44919, Republic of Korea*

I. Nam and G. Hahn

*Pohang Accelerator Laboratory, Pohang 37673, Republic of Korea*

An active plasma lens (APL) focuses the beam in both the horizontal and vertical planes simultaneously using a magnetic field generated by a discharge current through the plasma. A beam size of 5–10 $\mu$m can be achieved within a short distance using a focusing gradient on the order of 100 T/m. The APL is therefore an attractive element for plasma wakefield acceleration, because an ultrasmall size of the witness electron beam is required for injection into the plasma wakefield to minimize emittance growth and to enhance the capturing efficiency. When the drive beam and witness electron beam copropagate through the APL, interactions between the drive and witness beams, and the plasma must be considered. In this paper, through particle-in-cell simulations, we discuss the possibility of using an APL for the final focusing of the electron beam for the AWAKE RUN 2 experiments. It is confirmed that the amplitude of the plasma wakefield excited by proton bunches remains the same even after propagation through the APL. The emittance of the witness electron beam increases rapidly in the plasma density ramp regions of the lens. Nevertheless, when the witness electron beam has a charge of 100 pC, emittance of 10 mm mrad, and bunch length of 60 $\mu$m, emittance growth is not significant along the active plasma lens. For small emittance, such as 2 mm mrad, the emittance growth is found to be strongly dependent on the rms beam size, plasma density, and multiple Coulomb scattering.

## I. INTRODUCTION

An active plasma lens (APL) is a focusing element for charged particle beams. It focuses the beam in both horizontal and vertical planes using a magnetic field generated by the discharge current through a plasma. If we assume that the plasma and current density are uniform, the magnetic field and the associated focusing gradient can be derived from Ampere's law, as described in Eqs. (1)–(3) [1,2],

$$\frac{1}{r}\frac{\partial}{\partial r}(rB_\phi) = \mu_0 J_z, \tag{1}$$

$$B_\phi = \frac{\mu_0 J_z r}{2}, \tag{2}$$

$$g_r = \frac{\partial B_\phi}{\partial r} = \frac{\mu_0 I_0}{2\pi R^2}, \tag{3}$$

where $\mu_0$ is the permeability in vacuum, $J_z$ is the current density equal to $I_0/\pi R^2$, $I_0$ is the current induced in the APL, and $g_r$ is the focusing gradient. For instance, if the radius $R$ of the capillary plasma source (plasma radius) is 1.0 mm and the discharge current induced to the APL is 1 kA, then the focusing gradient is 300 T/m. This constant focusing gradient is the same for both the horizontal and vertical planes; the particle beam is focused in the radial direction by a single element.

---
[*]mchung@unist.ac.kr
[†]steffen.doebert@cern.ch

Beam focusing using an APL has been experimentally demonstrated [1,3–5]. In addition, it has been experimentally confirmed that abberation-free focusing can be achieved using argon plasma in the lens instead of lighter gases such as hydrogen or helium [2,6,7]. With lighter gases, the discharge current becomes nonuniform owing to heat transfer to the capillary wall [8], and this has been studied through MHD simulation and measurement of the discharge effect on electron probe beams [9,10]. When using a heavier gas, a uniform current density, and therefore a linear focusing gradient, can be maintained [11]. In addition, an analytical discussion of the transverse wakefield along the APL can be found in Ref. [12].

Considering the advantages of APL, it is expected that it can be used as a focusing element for injecting an electron beam into the plasma for wakefield acceleration, where an ultrasmall electron beam size is required to minimize emittance growth and to enhance the capturing efficiency. In particular, the advanced wakefield experiment (AWAKE) at CERN aims to demonstrate electron beam acceleration by a proton beam-driven plasma wakefield [13,14], as shown in Fig. 1. For the AWAKE RUN 2 experiments [15], two plasma sources are used. The first plasma source is used only for proton beam-seeded self-modulation (SSM) [16,17]. Once the proton beam is self-modulated, it is injected into the second plasma source for the acceleration stage. The electron beam is injected into the second plasma source using dipole magnets. In between the two plasma sources, a focusing element is installed for the electron beam.

It should be noted that the laser to ionize the vapor gas in the second plasma source is back-propagating against the direction of the witness electron beam [15,19]. The main reason of setting the laser direction in such a way is to shape an infinitely sharp-edged plasma density distribution at the entrance of the second plasma source, avoiding any effect from the plasma density ramp [15]. Therefore, even though it has been reported [20] that the beam size requirement could be relaxed by controlling the plasma density ramp at the second plasma source, we do not consider such a ramp in this study.

$$\sigma_r = \left( \frac{2c^2 \epsilon_n^2 m_e \epsilon_0}{n_p e^2 \gamma} \right)^{1/4}, \tag{4}$$

$$\beta_r = \left( \frac{2c^2 m_e \epsilon_0 \gamma}{n_p e^2} \right)^{1/2}, \qquad \alpha_r = 0. \tag{5}$$

Equations (4) and (5) [21] are the simple matching conditions for electron injection into the nonlinear plasma wakefield [22], where $e$ is the elementary charge, $c$ is the speed of light, $\epsilon_n$ is the electron beam normalized emittance, $m_e$ is the electron mass, $\epsilon_0$ is the vacuum permittivity, $n_p$ is the plasma density, and $\gamma$ is the relativistic factor of the beam. If the $\beta(=v/c)$ of the beam is close to one, the required Twiss betatron function $\beta_r$ in Eq. (5) is obtained from $\sigma_r = \sqrt{\beta_r \epsilon_{g,r}}$, where $\epsilon_{g,r}$ is the rms geometric emittance. From Eq. (4), when the normalized emittance of the electron beam is 2 or 10 mm mrad with an energy of 165 MeV, and the density of the plasma source for electron acceleration is $7 \times 10^{14}/\mathrm{cm}^3$, then the required rms beam size becomes 5.6 or 12.56 $\mu$m, respectively. In addition, the required Twiss $\beta$ function at the entrance of the plasma source for electron acceleration becomes 5.1 mm, which can be evaluated by Eq. (5).

A millimeter-scale of Twiss $\beta$ function is very tight to achieve by the conventional electron beamline using normal electromagnetic quadrupoles within a short distance. As shown in Fig. 1, the gap distance between two plasma sources should be less than 1.0 m to sustain the wake amplitude from the self-modulated proton bunches along the second plasma source [23]. Therefore, when the normal quadrupole magnets with at least three elements are used, the gap distance becomes longer than the requirement. A permanent-magnet quadrupole (PMQ) [24–26] may be considered as one of the options because its length can be made as short as 0.01 m and the maximum focusing gradient is in the order of sub kT/m. However, on-line adjustment of focusing strength of the PMQ during experiments is very challenging. Therefore, an APL can be an alternative and more attractive element for electron beam focusing.

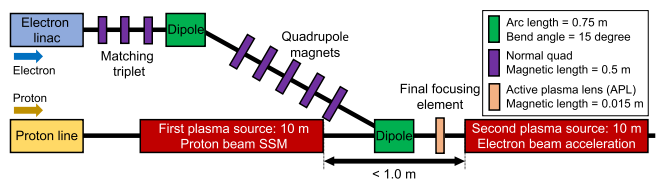

FIG. 1. Schematic view of the AWAKE RUN 2 experiment using two plasma sources (not to scale). The electron beam is generated by combining S- and X-band RF accelerators [18] and transported through the electron beamline. Normal quad indicates the normal electromagnetic quadrupole magnet. Note that the beamline in this schematic view, including the final focusing element, is not the baseline of the AWAKE experiment (i.e., the baseline does not consider using an APL).

In this study, we focus on the possibilities of using an APL for the final focusing of the witness electron beam for the AWAKE RUN 2 experiments. In Sec. II, the electron and proton beam parameters used for the simulation and the APL specifications are described. Using particle-in-cell (PIC) simulation, the focusing of the electron beam along the APL is discussed in Sec. III. Here, the interactions between the drive proton bunches, witness electron beam, and plasma in the APL are investigated. The emittance growth of the electron beam is also explored to verify whether the emittance growth is significant, and to examine the main factors of such growth along the APL. We also discuss the required discharge current of APL in Sec. V, as the operational limits of APL may be of concern for a real experiment. Regarding the issue of the discharge current, an attempt to lower the discharge current is discussed. Finally, the conclusion on whether the APL can be used for the final focusing of the witness electron beam for the proton beam-driven plasma wakefield acceleration is summarized in Sec. VI.

## II. SIMULATION PARAMETERS FOR THE APL STUDY

In this section, we present the initial electron beam parameters, APL parameters, and self-modulated proton bunches used in the simulation.

### A. Electron beam and APL parameters

The target electron beam parameters before entering the second plasma source (i.e., the acceleration column shown in Fig. 1) is summarized in Table I. The beam distribution was assumed to be Gaussian with zero energy chirp. The rms bunch length was set by following Ref. [27]. The target electron beam size was evaluated using Eq. (4), given the normalized emittance. The AWAKE baseline emittance out of the electron injector was 2 mm mrad. In this study, we used not only the baseline emittance, but also a larger value up to 10 mm mrad to parametrically study the evolution of the emittance along the APL, its growth ratio, and the variation of the slice emittance. The initial beam size before entering the APL was considered as a variable up to 250 $\mu$m. Furthermore, the Twiss $\alpha$ function of the initial electron beam was set to zero.

TABLE I. Electron beam parameters for the AWAKE RUN 2 experiment before injection into the plasma wakefield together with APL specifications. An emittance of 2 mm mrad was used as the baseline parameter.

| Parameter | Value |
| --- | --- |
| Energy, $E$ | 165 MeV |
| Energy spread, $\sigma_\delta$ | 0.1% |
| Charge, $Q$ | 100 pC |
| Bunch size, $\sigma_r$ | 5.6–12.6 $\mu$m |
| Bunch length, $\sigma_z$ | 60 $\mu$m (200 fs) |
| Normalized emittance, $\epsilon_n$ | 2–10 mm mrad |
| Radius of APL | 1.0 mm |
| Length of APL | 15 mm |
| APL plasma density, $n_p$ | $1.4583 \times 10^{17}/\mathrm{cm}^3$ |

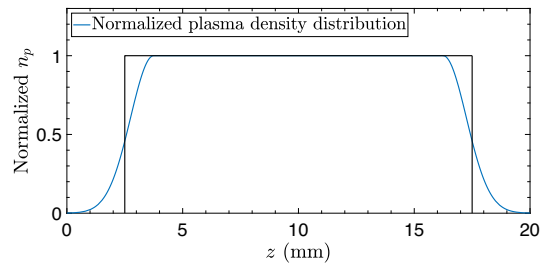

FIG. 2. Normalized plasma density distribution along the APL. The black solid line represents the active focusing gradient, which is assumed to be constant within the lens. We note that in the time scale of proton and electron beam passages, discharge current in the APL can be regarded as stationary [7].

The plasma density distribution along the APL is shown as a blue line in Fig. 2. It is assumed that the electrodes are placed at $z_1 = 2.5$ mm and $z_2 = 17.5$ mm. Therefore, the length of the APL column was set to 15 mm. The active focusing gradient inside this column was assumed to be constant. The plasma density ramp at both ends was set using a Gaussian distribution based on Ref. [28], which is described by Eq. (6).

$$n_p(z) = n_{p,0} \times \exp\left[-\left(\frac{z - z_{1,2}}{\sqrt{2}\sigma_{\mathrm{ramp}}}\right)^2\right], \qquad (6)$$

where $n_{p,0}$ is the peak plasma density of the APL, $\sigma_{\mathrm{ramp}}$ is the Gaussian ramp parameter, and $z_{1,2}$ are the start positions of the up-ramp and down-ramp regions. The value of $\sigma_{\mathrm{ramp}}$ estimated through measurements [28] is approximately 5.0 mm, but we used $\sigma_{\mathrm{ramp}}$ of 1.0 mm as a reference for this study. In addition, $z_1$ and $z_2$ are set to 3.75 mm and 16.25 mm, respectively. The peak plasma density along the APL is $1.4583 \times 10^{17}/\mathrm{cm}^3$, which corresponds to a pressure of 6 mbar at room temperature of 298 K. The radius of the APL is fixed to 1.0 mm unless otherwise mentioned.

### B. Setup of modulated proton bunches

For this simulation study, we generated the self-modulated proton bunches following Ref. [29] where the seeding method is ionization-front seeding [30]; only the back half of the proton bunches which interacted with the plasma were used. Front half of the proton beam was not considered in this APL simulation study.

Using the self-modulated proton bunches, we performed the PIC simulation using the FBPIC code [31] to verify the generation of a longitudinal plasma wakefield. Here, the plasma density was set to the value of the APL described in Table I. Figures 3(a) and 3(b) show the self-modulated proton bunches and the longitudinal wakefield in the APL plasma column. As in Fig. 3(b), the longitudinal wakefield is not constructive because the resonant conditions $k_p \sigma_r \leq 1$ and $k_p \sigma_z \leq \sqrt{2}$ are invalid for the given plasma density, rms beam size, and the length of each modulated proton bunch [32]. We note that the APL plasma density is much higher (about 1000 times) than that of the plasma source for the proton SSM. The longitudinal wakefield oscillates with a maximum wake amplitude of approximately 10 MV/m. On the left side of Fig. 3(b), some peaks and a gradual increase in the longitudinal wakefield are observed. This is associated with numerical noise, which is accumulated due to the long simulation box length [33,34]. An enlarged view of the longitudinal wakefield in the red box is shown in Fig. 3(c). The form of the longitudinal wakefield is sinusoidal. For the APL studies in the presence of the proton bunches, the simulation setup has been simplified due to (i) the issues of numerical noise in the case where many long proton bunches propagate to the short plasma column and (ii) the destructive nature of the wake amplitude. Therefore, we used only 11 modulated proton bunches as indicated in Fig. 4(a). These proton bunches are extracted from the originally self-modulated proton bunches [35] with finite extent $\Delta \xi = 15$ mm. An electron bunch is placed between the 10th and 11th proton bunches from the right, as shown in Fig. 4(b). The transverse density distribution of a single proton bunch is shown in Fig. 4(c), where the orange solid line represents the Gaussian fitting.

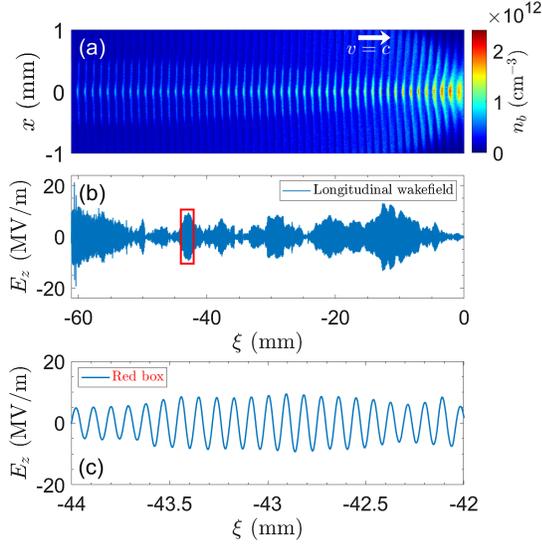

FIG. 3. Self-modulated proton bunches (a) and longitudinal wakefield (b) inside the APL plasma column with a density of $1.4583 \times 10^{17}/\text{cm}^3$. The propagation direction of the bunches is to the right. Subfigure (c) is the longitudinal wakefield inside the red box indicated in (b). The longitudinal coordinate $\xi = z - ct$.

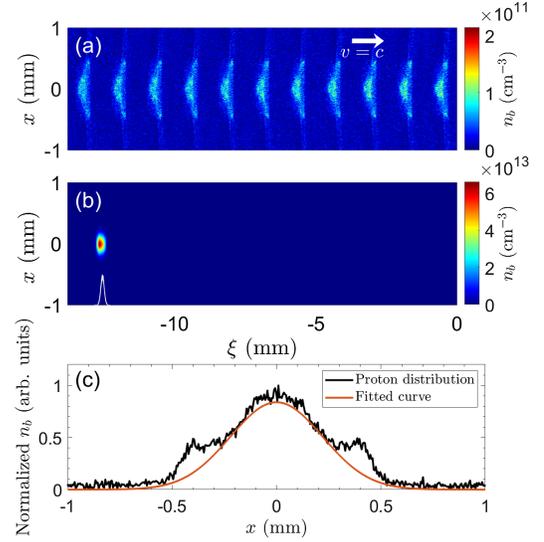

FIG. 4. (a) Eleven self-modulated proton bunches used for the APL simulation. (b) An electron bunch with an rms beam size of 100 $\mu$m was placed between the 10th and 11th proton bunches from the right. The propagation direction of the beam is toward the right. (c) Transverse density distribution of a single proton bunch.

The transverse rms beam size of the proton beam was 221 $\mu$m from the standard deviation obtained by Gaussian fitting. The calculated rms length of each proton bunch shown in Fig. 4(a) was approximately 107 $\mu$m.

## III. ELECTRON BEAM FOCUSING ALONG THE APL

In this section, we discuss the transverse focusing field generated by both the drive proton bunches and the electron beam, the evolution of the phase space, and the emittance of the electron beam during the propagation through the APL. For the PIC simulation setup, the transverse radius and the longitudinal length of the cylindrical simulation box were set to 1 and 20 mm, respectively. The resolution was set to 1/100 of the plasma wavelength, where the plasma density was $1.4583 \times 10^{17}/\text{cm}^3$. Then, the grid size of the simulation box was 0.87 $\mu$m in both the transverse and longitudinal planes. A convergence check was performed with a grid size of 0.58 $\mu$m, which corresponds to 1/150 of the plasma wavelength. The number of macroparticles for the electron beam used in the simulation was set to $1 \times 10^6$. To take into account the APL focusing gradient, an external magnetic field was used during the PIC simulation. Drive proton bunches are always considered together during the simulation unless otherwise mentioned.

### A. Plasma wakefield from the proton and electron beams

The longitudinal and transverse wakefields from the particle beam can be evaluated using Eqs. (7)–(9) [22].

$$W_z(\xi, r) = E_z(\xi, r)$$
$$= \frac{Ne}{\epsilon_0} \int_{-\infty}^{\xi} d\xi' n_{b\parallel}(\xi') \cos k_p(\xi - \xi') R(r), \quad (7)$$

$$W_r(\xi, r) = E_r(\xi, r) - cB_\theta(\xi, r)$$
$$= \frac{Ne}{\epsilon_0 k_p} \int_{-\infty}^{\xi} d\xi' n_{b\parallel}(\xi') \sin k_p(\xi - \xi') \frac{dR(r)}{dr}, \quad (8)$$

$$R(r) = k_p^2 \int_0^r r' dr' n_{b\perp}(r') I_0(k_p r') K_0(k_p r)$$
$$+ k_p^2 \int_r^{\infty} r' dr' n_{b\perp}(r') I_0(k_p r) K_0(k_p r'), \quad (9)$$

where $\xi$ is the longitudinal coordinate, $r$ is the radial coordinate, $n_{b\parallel}$ ($n_{b\perp}$) is the normalized longitudinal (transverse) beam density profile, $k_p$ is the plasma wavenumber, and $I_0$ and $K_0$ are the modified Bessel functions. Here, $R(r)$ and $\frac{dR(r)}{dr}$ determine the amplitude of the field component in the radial direction, and it is related to the transverse beam density. For a given beam charge, the peak beam density decreases as the rms beam size increases. Furthermore, it is expected that the maximum amplitude of $R(r)$ decreases as the beam density decreases.

First, Fig. 5 shows the maximum longitudinal plasma wakefield along the entrance ramp region of the APL calculated by using Eqs. (7)–(9). The longitudinal wakefield is maximized at the ramp region ($z = 0.7$ mm for proton beam and $z = 1.17$ mm for electron beam) where the resonant conditions of the linear plasma wakefield are satisfied. It should be noted that the plasma density at ramp region is still higher than those of the drive proton and witness electron beams used in this study. Therefore, it is expected that only the linear plasma wakefield is generated.

This feature of the plasma wakefield generated by the beams was confirmed with the PIC simulation results. Figure 6 shows the longitudinal and transverse components of the plasma wakefield along the APL. In this case, 4.9 [Figs. 6(a) and 6(b)] and 1.85 kA [Fig. 6(c)] of the discharge currents, which correspond to the APL focusing gradients of 979.5 and 369.7 T/m, were included in the simulations for the electron beam focusing. The longitudinal wakefield at the entrance of the APL is dramatically increased as expected from the theoretical estimation. In addition, it can be seen that the amplitude is maximized along the region where $1.0 \leq z \leq 1.3$ mm; this is the region where the plasma wakefield is generated by the witness electron beam as indicated in Fig. 5. As illustrated in Fig. 4, since the electron beam density is approximately hundred times larger than that of the proton bunch, the wake amplitude from the electron beam is dominant.

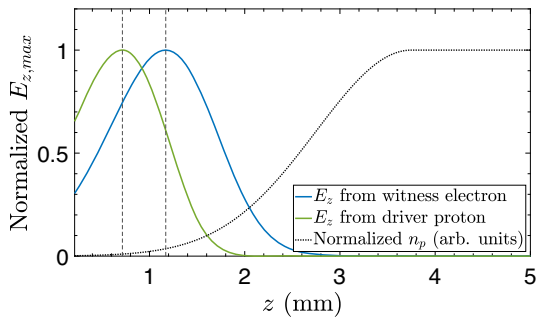

FIG. 5. Maximum normalized $E_z$ field generated by drive proton (green) and witness electron (blue) beams along the entrance region of the APL. The rms beam size and bunch length of the proton (electron) beam used for this calculation are 221(100) and 107(60) $\mu$m, respectively. Black dotted line indicates the APL plasma density distribution in arbitrary units.

In the region where the plasma density reaches a plateau of $1.4583 \times 10^{17}/\text{cm}^3$, the absolute value of the wake amplitude became a few MV/m, but gradually increased as the electron beam was focused and the beam density increased through the APL. In the second density ramp region, the longitudinal wakefield along the electron beam slice is much higher than that in the first ramp region because the increased electron beam density due to active focusing contributes to further development of the wake amplitude. It is of the order of 100 MV/m.

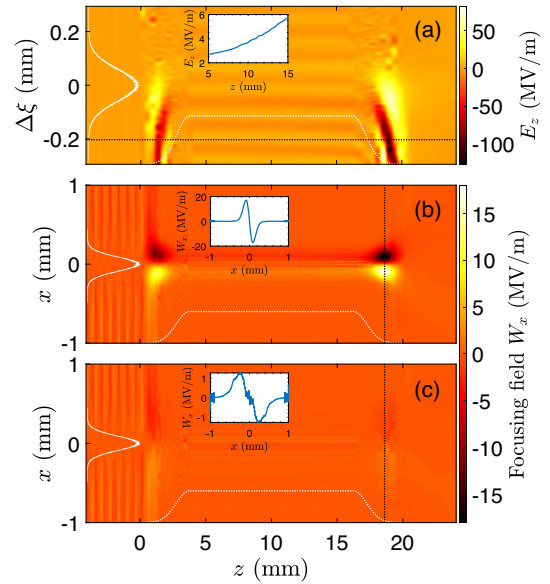

FIG. 6. Plasma wakefield obtained by the PIC simulation. (a) Longitudinal plasma wakefield $E_z$. The horizontal axis $z$ indicates the longitudinal coordinate of the APL, whereas the vertical axis $\Delta\xi$ represents the longitudinal coordinates of the electron beam slice. The electron beam head is placed at positive $\Delta\xi$, whereas the tail is located at negative $\Delta\xi$. Negative $z$ indicates the drift space before the APL. The solid and dashed white lines are the longitudinal electron beam distribution and plasma density distribution along the APL, respectively. Both distributions are given in arbitrary units. The inset in the figure shows the longitudinal wakefield at the black dotted line between 5 and 15 mm. (b–c) Horizontal plasma wakefield $W_x$ captured at the center of the electron beam longitudinal slice where the vertical axis $x$ represents the horizontal electron beam coordinates. The solid white line is the horizontal electron beam distribution. The insets in the figure show the horizontal wakefield at the black dotted lines. The initial electron beam sizes are (a,b) 100 and (c) 250 $\mu$m.

Likewise, the transverse plasma wakefield is large at both density ramp regions of the APL as described in Fig. 6(b). In this case, the initial electron beam size is 100 $\mu$m. This passive plasma wakefield is dramatically reduced when the beam with the larger initial beam size is injected as shown in Fig. 6(c), where the initial electron beam size is increased to 250 $\mu$m. For both cases, the focusing field is not linearly varied with respect to the transverse position. Therefore, it is expected that the transverse phase space of the electron beam will be nonlinearly distorted, leading to the increase of the phase space area.

## B. Evolution of the electron beam emittance

The evolutions of the electron beam emittances and final slice emittances are depicted in Fig. 7. It is noted that a slight difference in the initial emittance in each electron beam size case arises from the random generation of the initial electron beam distribution.

For all cases, the certain amount of the active focusing gradient is applied to focus the beam size down to the requirement evaluated by Eq. (4). The discharge current in the APL and the corresponding focusing gradient for each case of the initial electron beam size are listed in Table II. As illustrated in Figs. 7(a) and 7(b), the trend of the emittance growth is following that of the transverse wakefield shown in Fig. 6(b). Accordingly, at the exit plasma density ramp region, the emittance is largely increased compared to that along the region of the entrance density ramp. A gradual increase in the emittance can also be observed along the flat region of the APL because the increased electron beam density due to active focusing contributes to further development of the transverse plasma wake amplitude. This trend of the emittance growth is significantly reduced when the electron beam with the larger size is injected to the APL, because it leads to the reduction of the wake amplitude as described in Fig. 6(c). The normalized emittances after propagation through the APL with different initial electron beam sizes are summarized in Table III.

In the case where the initial electron beam emittance is 10 mm mrad, the emittance growth ratio was approximately 9.5% for the initial electron beam size of 100 $\mu$m. As the initial beam size increases, the growth ratio becomes less than 3%. The normalized emittance of the electron beam changes slightly when the plasma wakefield from proton bunches are considered as listed in Table III. Nevertheless, its contribution is negligibly small. Thus, we expect that the wakefield from the proton bunch does not significantly contribute to further degradation of the electron beam phase

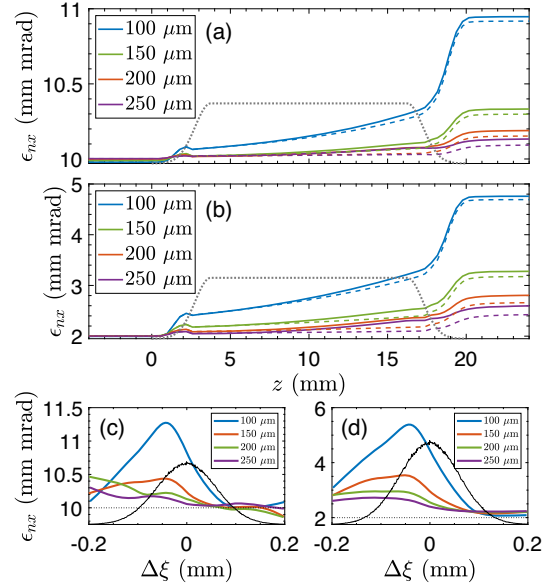

FIG. 7. (a,b) Evolutions of the normalized emittances along the APL with different initial electron beam sizes. The APL plasma (gray-dashed line in arbitrary units) is present between $z = 0$ and 20 mm, whereas the negative $z$ indicates the drift space through which the electron beam propagates. Solid lines indicate the emittance evolutions, including the wakefield effect from both the proton bunches and electrons, whereas the dashed lines represent the cases without proton bunches. (c–d) Slice emittances after propagation through the APL. Black line: Longitudinal electron beam distribution (arbitrary units), where the head is placed at positive $\Delta\xi$ and the tail is located at negative $\Delta\xi$. The initial emittances of the electron beams are (a,c) 10 and (b,d) 2 mm mrad, respectively.

space when the electron beam emittance is large enough, for example, 10 mm mrad.

In the case where the initial electron beam emittance is 2 mm mrad which is the AWAKE baseline parameter, however, the emittance growth is significant even when the initial electron beam size is 250 $\mu$m. The main reason for this emittance growth is that the effect of the transverse wakefield on the electron beam phase space is relatively large. Likewise, it can be seen that the wakefield from proton bunches further affects the electron beam emittance. The difference in emittance between the two cases listed in Table III increases as the initial electron beam size increases. The longitudinal wakefields along the electron beam slice at the plasma density ramp of the exit region are shown in Fig. 8. At negative $\Delta\xi$ (from $-0.2$ to $-0.1$ mm), the amplitudes of the wakefield including proton bunches are similar for both cases with different initial electron beam sizes. When the initial electron beam size is 100 $\mu$m, the total wakefield in this region is small compared to the maximum amplitude around $\Delta\xi = 0.07$ mm; its contribution to the increase in the projected emittance is relatively insignificant. However, the proton effect becomes dominant when the electron beam size is 250 $\mu$m.

TABLE II. Discharge current and focusing gradient of the APL required for beam size matching at the injection.

| Initial $\sigma_x$ ($\mu$m) | Initial $\epsilon_{nx}$ (mm mrad) | APL discharge current (kA) | Focusing gradient (T/m) |
|---|---|---|---|
| 100 | 10 | 4.90 | 979.50 |
|  | 2 | 3.04 | 607.00 |
| 150 | 10 | 3.15 | 629.09 |
|  | 2 | 1.71 | 342.14 |
| 200 | 10 | 2.32 | 463.54 |
|  | 2 | 1.18 | 237.29 |
| 250 | 10 | 1.85 | 369.73 |
|  | 2 | 0.95 | 190.38 |

TABLE III. Normalized emittance in horizontal plane after propagation through APL with different initial beam sizes. The vertical emittance is almost identical to that in the horizontal plane. $C_1$ is the case with considering the proton bunches and $C_2$ is the case without considering the proton bunches.

| Initial rms beam size ($\mu$m) | 100 | 150 | 200 | 250 |
|---|---|---|---|---|
| $\epsilon_{nx,i} = 10$ mm mrad | | | | |
| $C_1$ $\epsilon_{nx}$ (mm mrad) | 10.95 | 10.33 | 10.19 | 10.13 |
| $C_2$ $\epsilon_{nx}$ (mm mrad) | 10.92 | 10.30 | 10.15 | 10.09 |
| $\epsilon_{nx,i} = 2$ mm mrad | | | | |
| $C_1$ $\epsilon_{nx}$ (mm mrad) | 4.758 | 3.278 | 2.804 | 2.596 |
| $C_2$ $\epsilon_{nx}$ (mm mrad) | 4.692 | 3.174 | 2.659 | 2.423 |

Lastly, Figs. 7(c–d) show the slice emittances after propagation through the APL. As expected, different focusing fields are applied to each electron beam slice, resulting in a slice mismatch.

### C. Electron beam phase space distribution at focal point

Once the electron beam has propagated through the APL, the beam is focused, creating a beam waist at the focal point. To explore the phase space of the electron beam at the focal point, a particle tracking simulation was performed using the ASTRA code [36], which includes the space charge effect. Here, the final data obtained from the PIC simulation were considered as inputs for ASTRA tracking. Figure 9 shows the electron beam phase space at the focal point. The focused rms beam size and beam waist position after APL are listed in Table IV. The distance of the focal point is relative to the endpoint of the plasma channel, i.e., $z = 20$ mm. Each row in Fig. 9 indicates a different initial electron beam size from 100 to 250 $\mu$m. The first and second columns show the beam distribution and horizontal phase space.

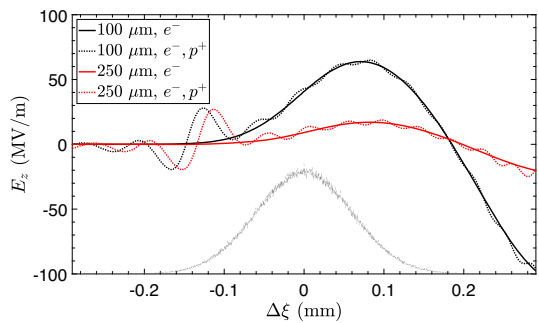

FIG. 8. Longitudinal wakefields at the exit plasma ramp region. Gray dashed line (arb. units) shows the longitudinal electron beam slice. The solid colored lines indicate the wakefields from the electron beam only, whereas the dotted lines represent cases with proton bunches. The initial electron beam sizes are 100 and 250 $\mu$m.

When the initial electron beam size is 100 $\mu$m, the horizontal phase space is distorted owing to the strong nonlinear transverse wakefield applied along the electron beam slice, as shown in Figs. 9(b) and 9(f). Particularly, as the emittance is significantly increased in the case where the initial emittance is 2 mm mrad, the distortion of the phase space from the nonlinear passive effect becomes more noticeable. As the initial beam size increases, the nonlinearity of the horizontal phase space decreases, as shown in Figs. 9(d) and 9(h).

For the rms energy spread, we found from the simulation that it remained the same as the initial value of 0.1% even though the longitudinal phase space is slightly distorted owing to the existing longitudinal wakefield. According to the values in Table IV, the beam sizes at the focal point, as illustrated in Figs. 9(a), 9(c), 9(e), and 9(g), almost satisfy the beam size requirement determined by Eq. (4). Fine adjustments of the APL discharge current and position would control the exact electron beam size.

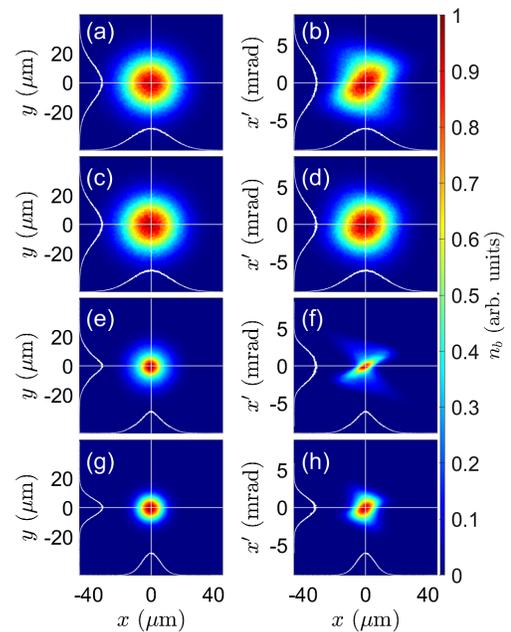

FIG. 9. Electron beam distributions (first column) and horizontal phase spaces (second column) at the focal point. (a–b) Initial beam size of 100 $\mu$m. (c–d) Initial beam size of 250 $\mu$m. (a–d) Initial emittance is 10 mm mrad. (e–f) Initial beam size of 100 $\mu$m. (g–h) Initial beam size of 250 $\mu$m. (e–h) Initial emittance is 2 mm mrad.

TABLE IV. RMS beam size at the focal point and beam waist position after the APL with different initial beam sizes. The required $\sigma_x$ was calculated using Eq. (4) with the final normalized emittance listed in Table III.

| Initial rms beam size ($\mu$m) | 100 | 150 | 200 | 250 |
|---|---|---|---|---|
| $\epsilon_{nx,i} = 10$ mmrad | | | | |
| Required $\sigma_x$ from Eq. (4) ($\mu$m) | 13.13 | 12.77 | 12.68 | 12.64 |
| Focused $\sigma_x$ ($\mu$m) | 12.74 | 12.68 | 12.72 | 12.69 |
| Beam waist position (mm) | 27.2 | 49.3 | 70.7 | 91.1 |
| $\epsilon_{nx,i} = 2$ mmrad | | | | |
| Required $\sigma_x$ from Eq. (4) ($\mu$m) | 8.67 | 7.19 | 6.63 | 6.40 |
| Focused $\sigma_x$ ($\mu$m) | 8.46 | 7.12 | 6.68 | 6.21 |
| Beam waist position (mm) | 47.3 | 95.5 | 144.0 | 183.1 |

Additionally, because the electron beam is rapidly focused downstream of the APL, the beam waist position is placed near the APL, reducing the drift space to less than 0.2 m. The gap distance between the two plasma sources can therefore be shortened when using the APL, sustaining the amplitude of the proton beam-driven plasma wakefield along the second plasma source [23].

### D. Effect of nonlinear focusing field from discharge current

Previous discussion of the electron beam focusing was based on the assumption that the magnetic field induced by the discharge current is linear. To discuss the further evolution of the emittance by the nonlinear focusing field from the discharge current reported in Refs. [9,10], we performed additional PIC simulations. Figure 10 shows the evolution of the emittance along the APL and horizontal phase space after the APL.

Through Fig. 10(a), one can see that the electron beam with a larger beam size such as 250 $\mu$m is more widely affected by the nonlinear focusing field, whereas the nonlinear effect is minimized as the initial beam size becomes small. Thus, as shown in Figs. 10(b–c), in the case where the initial electron beam size and emittance are 250 $\mu$m and 2(10) mm mrad, the final emittance after propagation to the APL becomes 6.22(14.24) mm mrad. The emittance growth ratio is then 140(38)% compared to the final emittance value with the linear focusing field only. One can also see that the emittance is gradually increasing along the flat region of the APL where the nonlinear focusing field from the discharge current is applied to the beam. However, when the initial electron beam size is 100 $\mu$m, the emittance growth ratio becomes 6.2(1.8)%, which is significantly reduced from the larger beam size case.

Additionally, as shown in Figs. 10(d–e), the phase space (orange scatter) is nonlinearly rotated following the profile of the focusing field, leading to the emittance increase. Therefore, when the nonlinear focusing field exists in the APL, the beam size should be properly chosen to minimize the effects of both the nonlinear focusing field of the APL and transverse wakefield from the beam, as experimentally discussed in Ref. [5].

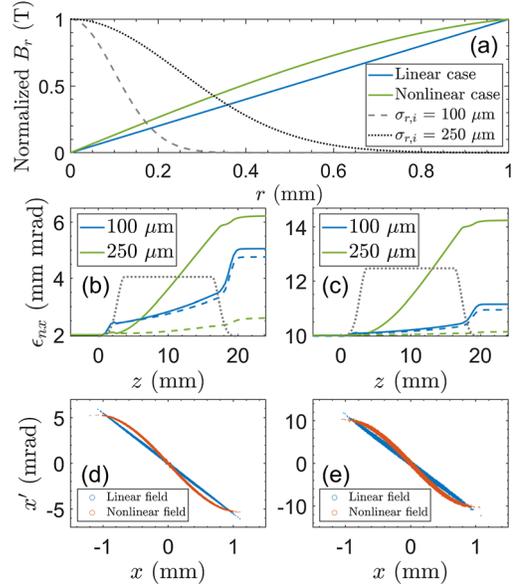

FIG. 10. (a) Focusing fields generated by the discharge current. (b–c) Evolutions of the emittance along the APL. Dashed lines indicate the cases with linear focusing field only whereas the solid lines represent the cases with nonlinear field. (d–e) Horizontal phase spaces right after the APL with the initial beam size of 250 $\mu$m. The initial emittances are (b,d) 2 mm mrad and (c,e) 10 mm mrad.

### E. Effect of proton beam modulation and defocusing in the APL

We also investigated whether proton bunches are defocused considerably by the active focusing gradient and transverse wakefield in APL. This could decrease the proton beam density, reducing the wake amplitude along the second plasma source. To verify the amplitude of proton beam-driven plasma wakefield, simulations were performed for the following two cases: (i) a reference case where proton bunches pass through only a drift and then enter the second plasma source; (ii) proton bunches pass through the APL and then enter the second plasma source. In the reference case, the total length of the APL was considered as an additional drift space. In both cases, the plasma density of the second plasma source was fixed to $7 \times 10^{14}$/cm$^3$ and the entrance of the plasma source was located at the electron beam waist position, as listed in Table IV. An APL discharge current and focusing gradient in Table II were used in this simulation. It was confirmed by the PIC simulation that the longitudinal wakefield created by the proton beam after passing through the APL is almost identical to the case without an APL element. Therefore, we can expect that the proton bunches are not significantly modulated inside a short APL plasma column.

## IV. PARAMETER SCAN

In the AWAKE experiment, it is important to maintain the emittance of the electron beam as close as the value just before entering the APL. Therefore, parameter scanning was performed to find the optimal conditions of the injected electron beam and APL, where the emittance growth induced by the passive wakefield is minimized.

First, we performed the PIC simulations by changing the electron beam charge and bunch length while keeping the peak current of the beam to the value calculated by the parameters indicated in Table I. Initial electron beam size is 100 $\mu$m. Figure 11 shows the final emittance and transverse wakefield $W_r$. As shown in Fig. 11(a), in the case of a 10 pC beam charge, the emittance was 2.19 mrad for both cases with and without ramp. As the electron beam charge increases, the final emittance for the case with the plasma density ramp also increases. It is due to the increased passive wakefield along the electron beam, as described in Fig. 11(b) which shows the maximum wake amplitude along the ramp using Eqs. (8)–(9) with considering only the electron beam. In this calculation, we fixed a certain position in the ramp region with plasma density $n_p = 1 \times 10^{16}/\mathrm{cm}^3$, which is approximately 10% compared to the peak APL plasma density. Through Figs. 11(a) and 11(b), one can see that the trend of the wake amplitude in the transverse plane is following the trend of the final emittance. When the bunch length is fixed, the wake amplitude is expected to be linearly increased as the beam charge increases, which can be deduced from Eq. (8). However, when bunch length is also varied to keep the peak current constant, the increase of the wake amplitude is not only related to the charge but also to the resonant condition of the linear plasma wakefield; $k_p \sigma_z$ approaches to $\sqrt{2}$ as the charge increases, leading to maximizing the wake amplitude.

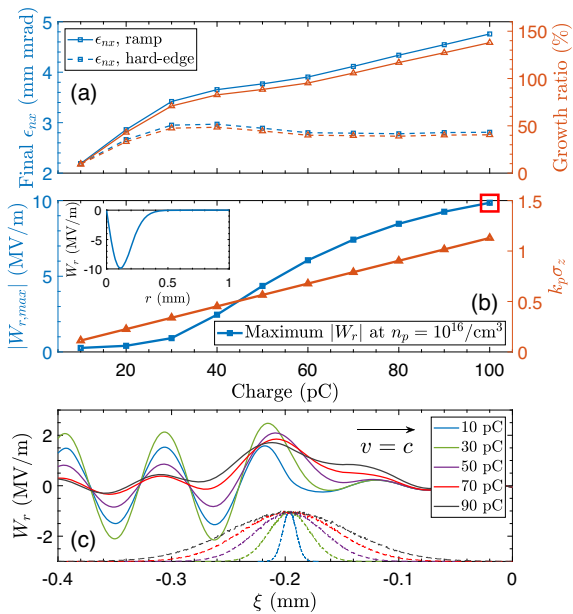

FIG. 11. (a) Final emittance and its growth ratio as a function of electron beam charge. The bunch length was varied to maintain the peak current same for all cases. Solid and dashed lines indicate the cases with and without the plasma density ramp, respectively. (b) Maximum $|W_r|$ value (blue curve) at the center of the electron beam slice obtained by Eq. (8) with plasma density $n_p = 1 \times 10^{16}/\mathrm{cm}^3$. Inset shows the transverse wakefield along $r$ in the case of 100 pC charge (red box). In this theoretical calculation, only the electron beam is considered. Orange curve shows the resonant condition $k_p \sigma_z$. (c) Transverse wakefields along the electron slice distribution with different charges obtained by the PIC simulation. Dashed Gaussian histograms are the longitudinal beam distributions. Here, the electron beam is moving through the flat region of the APL.

Meanwhile, when the plasma density of the APL is uniform (hard-edged), behavior of the final emittance [dashed lines in Fig. 11(a)] is somewhat different compared to that with the density ramp. The final emittance is maximized at the beam charge of 30 pC, and becomes constant as the beam charge increases. This can be understood by examining the transverse wakefield as illustrated in Fig. 11(c). When the beam charge is 30 pC, the wake amplitude along the electron beam slice is maximized compared to the other cases. However, as the beam charge increases further, the wake amplitude decreases and becomes saturated; the amplitude of the wakefield from the proton bunches is now dominant. This trend is illuminated on the emittance variation as a function of the beam charge. In the case where the beam charge is 100 pC, the final emittance and its growth ratio with the hard-edged APL become 2.81 mm mrad and 40%, respectively. Compared to the case with the plasma density ramp, the emittance growth is significantly reduced. Therefore, in order to reduce the emittance growth, the plasma density ramp should be properly controlled so that the effect of the passive wakefield is minimized.

To further discuss the minimization of the emittance growth, we performed simulations using an increased plasma density and verified whether the final emittance could be reduced. Here, the plasma density of the APL was changed from $1.4583 \times 10^{17}$ to $3 \times 10^{17}$ and $5 \times 10^{17}/\mathrm{cm}^3$. The plasma density ramp was considered at both ends. The initial electron beam charge and emittance were fixed at 100 pC and 2 mm mrad, respectively. According to Ref. [37], an increased plasma density decreases the wake amplitude. Consequently, the final emittance after propagation through the APL is decreased. Figure 12 shows the normalized electron beam emittance with different initial electron beam sizes and peak APL plasma densities. When the initial electron beam size was 250 $\mu$m and the APL plasma densities were $3 \times 10^{17}/cm^3$ and $5 \times 10^{17}/cm^3$, the final emittance (growth ratio) became 2.494 mm mrad (24.7%) and 2.452 mm mrad (22.6%), respectively. Because these values are obtained by considering the APL plasma density ramp, we expect that the emittance growth could be further reduced by optimizing the length and/or shape of the ramp region.

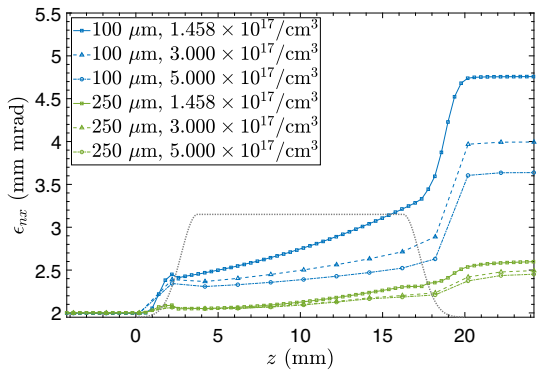

FIG. 12. Evolution of emittance along the APL with several different peak APL plasma densities. The grey-dashed line represents the APL plasma distribution in arbitrary units. The initial electron beam sizes used for this simulation were 100 and 250 $\mu m$.

This trend, however, is the result in which the effect of multiple Coulomb scattering (MCS) was not taken into account. Even though the emittance growth due to the MCS was not noticeable in the APL experiments reported in Refs. [2,6] where the pressure was 6 mbar, in general MCS can alter the phase space as the pressure and Twiss beta function increase further [38]. Therefore, it is necessary to consider the corresponding MCS to more precisely evaluate the emittance variation (see Appendix for detailed discussion). In addition, to reduce the effect of aberration and MCS simultaneously, it is also considered to use neon gas alternatively [2,6] where the atomic number is lower than that of the argon gas. Since the ions are practically immobile in the PIC simulations, the results presented for argon gas so far remain nearly unchanged for neon.

## V. POSSIBILITIES FOR REDUCTION OF THE APL DISCHARGE CURRENT

In the previous sections, we discussed the focusing of an electron beam through the discharge current in an APL. For all cases with various initial electron beam sizes, high discharge currents are required. The discharge current in the order of kA is rather large compared to the other cases reported in Ref. [12], and it is expected to be somewhat challenging to generate for the long-term operation due to certain engineering issues.

Therefore, to reduce the discharge current, use of a smaller aperture size such as the plasma radius of 0.5 mm was considered. We found from the PIC simulation that the amplitude of the longitudinal wakefield with a reduced aperture size is almost identical to that of the reference case since the focused proton bunches from the SSM are concentrated within $r = 0.5$ mm, as shown in Fig. 4.

However, because the protons defocused by the SSM [29,39] hit the APL aperture in this scheme, another simulation was performed to estimate the proton beam energy deposited to the APL aperture, verifying whether the defocused proton particles by the SSM are a possible source of damage to the APL. This was done using GEANT4 [40,41] with the FTFP_BERT_EMV physics list through the MiniScatter application [38]. In this simulation, we considered the worst case where the entire proton beam ($3 \times 10^{11}$ particles) hits a 15 mm long sapphire APL. Even in that case, the estimated energy deposited onto the APL was 1.2 J. Two 5 mm copper electrodes on each side increased the energy deposition in the sapphire to 1.4 J. Even though a study of the implications of generated particle showers to the radiation environment in the tunnel may be necessary, the energy deposition from the proton beam onto the APL does not appear to be a significant issue.

By using a reduced radius for the APL, the maximum acceptable initial rms size of the electron beam before entering the APL was reduced to 150 $\mu m$. When the initial emittance of the electron beam was 2 mm mrad, the required discharge current became 758.78 and 427.67 A for initial beam sizes of 100 and 150 $\mu m$, respectively. Therefore, if the initial emittance is small enough, practical application of the APL is possible in the context of the required discharge current. Likewise, when the initial emittance is changed to 10 mm mrad, the required discharge current becomes 1.225 kA and 780 A. Even though the discharge current is significantly reduced from the values indicated in Table II, the long-term stability of the lens in terms of the generation of the discharge current and damage of the APL must be verified in all cases.

## VI. CONCLUSION

In this study, we showed that focusing of an electron beam using an APL in the presence of drive bunches is feasible. We also demonstrated that the requirements for witness beam injection into a plasma for wakefield acceleration can be achieved by careful choice of input beam and APL parameters. This enables the use of a compact focusing device that allows for short drift distances, which is highly beneficial for maintaining the required quality of the drive beam. We discussed the phase space of the electron beam and its evolution along the APL using PIC simulations. It was found that the distortion of the transverse phase space is significant in the region of the plasma density ramp. This leads to a mismatch in the slice distribution, resulting in projected emittance growth. Therefore, to minimize the emittance growth along the APL, the plasma ramp at both ends should be as sharp as possible.

Because the main source of the emittance growth is from the plasma density ramp along the APL, more study into the ramp region and its impact on beam emittance is required. In addition, since the current density may be distorted by the modulated plasma electrons from the drive and witness beams, the magnetic field generated by the discharge current could also be perturbed accordingly. Such an effect needs to be explored further.

For long term, stable operation of the APL, reduction of discharge current could be advantageous. A possible option is to use a smaller aperture size of the APL, for example, $R = 0.5$ mm. PIC simulation results verified that the use of a small aperture-size APL will not significantly affect the quality of proton bunches. Moreover, it was confirmed by the GEANT4 simulation that the energy deposited to the APL will be of the order of 1 J. Therefore, the damage on the APL due to defocused proton beam particles may not be a significant issue.

With an initial electron beam charge of 100 pC and emittance of 10 mm mrad, the emittance growth after propagation through the APL turned out to be insignificant. This result may be further improved when optimizing the length and shape of the density ramp region.

Finally, we confirmed that modulated proton bunches were not significantly disturbed by the APL, and that the generation of a longitudinal wakefield in the second plasma source was not affected by the insertion of the plasma lens.

## ACKNOWLEDGMENTS

This work was supported by the National Research Foundation (NRF) of Korea (Grant No. NRF-2016-Global Ph.D. Fellowship Program and NRF-2020R1A2C1010835). This research was also supported by the R&D program (code No. IN2004-6) through the Korea Institute of Fusion Energy, funded by government funds. This work was also supported by the Research Council of Norway (NFR Grant No. 310713).

## APPENDIX: ESTIMATION OF MULTIPLE COULOMB SCATTERING EFFECTS

Since the transverse rms emittance is $\epsilon_\perp (= \epsilon_x = \epsilon_y) = \sqrt{\langle x^2 \rangle \langle x'^2 \rangle} = \langle x^2 \rangle \beta_\perp$ at a beam waist, the increase in the rms emittance during the transit of $\Delta z$ along the scattering medium is approximated as [42]

$$\Delta \epsilon_\perp \simeq \beta_\perp \Delta \langle x'^2 \rangle \simeq \beta_\perp \theta_{\text{rms},x}^2. \quad (A1)$$

The normalized emittance $\epsilon_n$ is given by $\epsilon_n = \epsilon_\perp (\beta\gamma)$. Therefore, we have the rate of the change of the normalized emittance as [42]

$$\frac{\Delta \epsilon_n}{\Delta z} \simeq (\beta\gamma)\beta_\perp \frac{\theta_{\text{rms},x}^2}{\Delta z} \simeq (\beta\gamma)\beta_\perp \frac{d\theta_{\text{rms},x}^2}{dz}. \quad (A2)$$

We note that emittance increase from multiple-scattering depends on the transverse beta function $\beta_\perp$ at the medium.

First, let us consider the scattering through neutral vapor with density $n$. The rate of rms scattering angle of an ultrarelativistic ($\beta \simeq 1$) electron through a neutral vapor can be given in terms of the radiation length $L_R$ [43]:

$$\left(\frac{d\theta_{\text{rms},x}^2}{dz}\right)_{\text{vapor}} = \frac{1}{2}\left(\frac{d\theta_{\text{rms}}^2}{dz}\right)_{\text{vapor}} = \frac{1}{L_R}\left(\frac{E_s/\sqrt{2}}{\gamma m_e c^2}\right)^2, \quad (A3)$$

where $E_s$ is the characteristic scattering energy ($\approx 20$ MeV to a good approximation). The radiation length is given by [43]

$$L_R = \frac{N_A \times 716.4 \ [\text{cm}^{-2}]}{n[\text{cm}^{-3}] \times Z(Z+1)\ln(287/\sqrt{Z})}, \quad (A4)$$

where $Z$ is the atomic number of the medium (18 for argon gas and 10 for neon gas) and $N_A$ is the Avogadro's number.

Second, we can estimate the scattering from an ion column as follows [43]:

$$\left(\frac{d\theta_{\text{rms},x}^2}{dz}\right)_{\text{ion}} = \frac{k_p^2 r_c}{\gamma^2} Q_i \ln \Lambda, \quad (A5)$$

where the plasma wavenumber $k_p = \sqrt{n_p e^2/m_e \epsilon_0 c^2}$ (note that there is a typo on the definition of $k_p$ in Ref. [43]), $n_p = Q_i n$ is the plasma density, $Q_i$ is the ion charge number, and $r_c$ is the classical electron radius. The Coulomb logarithm $\Lambda$ is approximately in the order of 10 for typical laboratory plasmas.

Indeed, the equation for the rate of change of the normalized emittance has an energy loss cooling term as well [42].

$$\frac{\Delta \epsilon_n}{\Delta z} \simeq -\frac{1}{(\gamma m_e c^2)} \rho \left(\frac{dE}{dx}\right) \epsilon_n, \quad (A6)$$

where $\rho$ is the mass density of the medium and $(dE/dx)$ is the stopping power. For the practical parameters of the APL considered here, nevertheless, the energy loss cooling term is negligible.

For the base parameters used in the main article [$n = 1.4583 \times 10^{17}/\text{cm}^3$, plasma length $\Delta z = 15$ mm, electron beam energy $(\gamma - 1)m_e c^2 = 165$ MeV, and $Q_i = 1$], we calculate the relative changes in normalized emittance as summarized in Table V.

Depending on the values of the initial normalized emittance and rms beam size, multiple scattering effects can be competing with the other effects introduced in the main article. To minimize the scattering effects, we need to have a low $\beta_\perp = \langle x^2 \rangle / \epsilon_\perp$ at the APL. For the cases of

TABLE V. Emittance change by multiple Coulomb scattering.

| Initial $\epsilon_{nx}$ (mm mrad) | 2 | 2 | 10 | 10 |
|---|---|---|---|---|
| Initial $\sigma_x$ ($\mu$m) | 100 | 250 | 100 | 250 |
| $\left(\frac{\Delta\epsilon_n}{\epsilon_n}\right)_{vapor}$ (%) with $Z = 18$ | 139.5 | 871.9 | 5.58 | 34.9 |
| $\left(\frac{\Delta\epsilon_n}{\epsilon_n}\right)_{ion}$ (%) with $Z = 18$ | 0.544 | 3.4 | 0.021 | 0.136 |
| $\left(\frac{\Delta\epsilon_n}{\epsilon_n}\right)_{vapor}$ (%) with $Z = 10$ | 48.0 | 300 | 1.92 | 12.0 |
| $\left(\frac{\Delta\epsilon_n}{\epsilon_n}\right)_{ion}$ (%) with $Z = 10$ | 0.544 | 3.4 | 0.021 | 0.136 |

250 $\mu$m beam size, therefore, the multiple scattering effects are dominant and become the major source of emittance growth. The multiple scattering effects become more serious for higher plasma density and atomic number, thus there is a certain limit in choosing the plasma density and atomic number. On the other hand, use of a neon gas can reduce the multiple scattering effects. In many cases, we expect that the effects of multiple Coulomb scattering and plasma ramp are equally important in determining emittance variation along the APL.

———————————